\documentclass[11pt,twocolumn]{article}

\usepackage{times}
\usepackage{helvet}

\usepackage[margin=1in]{geometry}

\usepackage[small,compact]{titlesec}
\titleformat*{\paragraph}{\sffamily\bfseries\small}

\usepackage[para]{footmisc}

\usepackage{fancyhdr}
\begin{document}

\thispagestyle{plain}

\begin{center}
{\sffamily \bfseries Modeling Astrophysical Explosions with Sustained 
                     Exascale Computing\footnotemark[1]} \\
{\small
M.~Zingale\footnotemark[2]\footnotemark[5], 
A.~C.~Calder\footnotemark[2], 
C.~M.~Malone\footnotemark[3],
\& F.~X.~Timmes\footnotemark[4]
}
\end{center}
\footnotetext[1]{Response to RFI NOT-GM-15-122: {\em Science Drivers Requiring Capable Exascale High Performance Computing}}
\footnotetext[2]{Stony Brook University}
\footnotetext[3]{Los Alamos National Laboratory}
\footnotetext[4]{Arizona State University}
\footnotetext[5]{michael.zingale@stonybrook.edu}

Our understanding of stars and their fates is based on coupling
observations to theoretical models.  Unlike laboratory physicists, we
cannot perform experiments on stars, but rather must
patiently take what nature allows us to observe.  Simulation offers a means
of virtual experimentation, enabling a detailed understanding of the
most violent ongoing explosions in the Universe---the deaths of stars.

Stars can explode in a surprising variety of ways, driven by either
nuclear or gravitational potential energy release.  The
explosion can consume the entire star (or stars) or just
surface layers, and exotic remnants, like neutron stars or black holes
can be left behind.  Stellar explosions are critically important sources of
nucleosynthesis, and enrich the interstellar medium with heavy elements.
All of the iron in the Universe, for example, was synthesized in
stellar explosions.

Both the DOE (including the national labs) and the
NSF have supported the development and application of simulation codes
to stellar explosions.  Many successes have been met, but there are
still great uncertainties in the mechanisms of core collapse
supernovae (the death of massive stars) and thermonuclear supernovae
(the explosion of compact white dwarf stars).  Astronomy is increasingly
in the ``big data'' era, with survey telescopes
like LSST coming online toward the end of the decade that will greatly
expand observations of astrophysical transients, likely finding entire
new classes to be understood.

Cutting edge research in stellar astrophysics is performed with both
large, multidimensional simulations, demanding 10s of millions of
core-hours, as well as suites of one-dimensional evolutionary
simulations with exceptionally detailed microphysics.  The interplay
between these two paths is critical to building a physical picture
of stellar explosions, and each has unique (and increasing)
computational demands.  For brevity, here we focus primarily on the
needs for multi-dimensional work.

The standard practice in stellar astrophysics is to describe the star
as a fluid and use domain decomposition to divide the work across
computational nodes.  Ideally, shared memory parallelism is used
within a node---reducing the memory overhead---with message
passing used across nodes.  Work on exploiting accelerators (GPUs and
Intel Xeon Phi processors) is underway for many codes, and standard
technologies (OpenMP, OpenACC, and MPI) allow for portability.

An example of a success in our field is enormous progress made over
the last decades in understanding thermonuclear (type Ia) supernovae.
Core models were developed in the 1990s through one-dimensional
calculations, raising a host of complex questions about the physical
processes in the stars.  Large-scale multi-dimensional simulations
followed these and explored a variety of progenitor systems and explosion
mechanisms, allowing researchers to address not just the question of a
successful explosion, but deeper issues such as systematic effects on
the brightness of an event and explanations for unusual or outlying
events.  Recently, three-dimensional progenitor models (of both types
of supernovae) became feasible, greatly increasing our understanding
of the initial conditions of the explosion.  These problems require
the interaction of many different researchers and simulation codes
capable of modeling the different phases.

\paragraph*{Increasing computing power.}

An increase of 100$\times$ in computing power will allow for
simulations at unprecedented fidelity. Fluid flows are chaotic, and a
range of instabilities and turbulence are ever present in models of
exploding stars. The goal of extant simulations is to understand the
feasibility of different theoretical models for explosions and to
probe the physics of the explosion mechanism itself.  It was only
recently that simulations switched from being predominantly
two-dimensional to three-dimensional---enabled by the large increase
in computing power over the last decade.  However, the range of length
and time scales that can be captured through simulation is a small
fraction of the true scales in stars.  This means that approximations
are made either explicitly (by introducing subgrid scale
models) or implicitly (having a numerical dissipation that operates on
much large scales than nature would have).  Convergence studies
(changing the resolution and seeking the same qualitative behavior)
can test our assumptions, but the looming question is whether there is
some resolution, yet unobtainable, where the qualitative nature of the
solution will change.  The promised increase in computing power will
allow for a much greater range of length scales to be modeled.

The increase in computer time also will allow for an expansion of the
physics modeled.  Current multi-dimensional simulations use small
nuclear reaction networks ($\sim$10--20 nuclei), which approximately
capture the energetics of the flow, but are unable to make detailed
predictions of the nucleosynthetic yields.  Additional pieces, like
radiation transport are crudely approximated (if modeled at all).  The
increase in computational power will allow for more realistic physics
throughout.

Finally, such an increase in computing time would bring today's
leadership computing into the realm of routine, allowing investigators
to perform many times the present volume of production simulations.
In astrophysics, this increase will allow suites of simulations to
investigate the sensitivity of events on both model and physical
parameters, enabling formal uncertainty quantification at
unprecedented levels.  This is true of both the multi-dimensional and
one-dimensional simulations.

\paragraph*{Impacts.}
In addition to answering astrophysical questions, research into
modeling astrophysical explosions at the exascale will significantly
impact many fields of science that address multi-scale, multi-physics
applications.  The simulation codes developed for these problems have
application to terrestrial combustion phenomena, climate and
atmosphere models, and DOE laboratory interests.  Importantly, this
research provides an excellent training ground for the next generation
of computational scientists, who can take their training to other
disciplines (and industry).

\paragraph*{Capabilities needed.}  Computational fluid dynamics requires
high performance interconnects, as nearest neighbor (and global for
some physics) communication is needed each timestep.  This requires
supercomputers instead of simple clusters.  A major challenge with the
increase in simulation size is the analysis of data (100s of TB
per simulation).  In situ data analysis will need more emphasis in the
future---many of the pieces for this are coming into place.

Supercomputing centers often favor ``hero'' calculations---those that
use a significant fraction of the machine.
However, science often needs capacity computing---many parameter
studies of the system help us understand the robustness of our models.
Going forward, there is a need for both capacity and leadership-class
computing centers.

Finally, there is an increasing desire to share simulation results,
which may be the seeds for follow-on simulations.  This
requires guaranteed long-term storage accessible to the community as a
whole.

\paragraph*{Foundational issues.}

The progress in our field is driven not just by increased FLOPS, but
also through algorithmic innovations.  Developing, maintaining, and
supporting simulation codes takes considerable effort, crossing
interdisciplinary lines (with coordination between domain scientists,
mathematicians, and computational scientists).  Funding mechanisms
need to recognize and support interdisciplinary work (for
example, as with the SciDAC program).

A further issue is that
often code development work is not given the same recognition and
rewards as the scientific results themselves.  This puts the code
developers, especially those early in their careers, at a competitive
disadvantage.  Perceptions of the role of code work will need to
adapt.  Likewise, increased support for code development and community
support through the traditional grant process would greatly help to
capitalize on code investments.  Open source codes also greatly help
amortize the costs of code development, and enable (and encourage)
reproducibility of results, a hallmark of science.  The astrophysics
community does a reasonable job in making codes available, and
incentive structures should be setup to further encourage this.

Awards of computer time don't come with monetary support for the
researchers who will run and analyze the simulations, and grants don't
come with a guarantee of computer time.  This is something of a
``chicken-and-egg" problem, as it is necessary to have both in place
independently to make effective use of either resource.  Mechanisms
linking the two should be developed.

Finally, continued support for training of students is essential. The
Argonne Training Program on Extreme-Scale Computing is an excellent
example.

\end{document}